\documentclass[twocolumn]{aastex631}

\usepackage[version=3]{mhchem}
\usepackage{hhline}
\usepackage[shortlabels]{enumitem}

\newcommand{\ms}{\ensuremath{\rm m\,s^{-1}}}
\newcommand{\kms}{\ensuremath{\rm km\,s^{-1}}}
\newcommand{\gs}{\ensuremath{\rm g\,s^{-1}}}
\newcommand{\gcc}{\ensuremath{\rm g\,cm^{-3}}}
\newcommand{\teff}{\ensuremath{T_{\rm eff}}}
\newcommand{\logg}{\ensuremath{\log{g}}}
\newcommand{\vsini}{\ensuremath{v \sin{I_\star}}}

\shorttitle{Metals and potential escape in HAT-P-67b}
\shortauthors{Bello-Arufe et al.}
\graphicspath{{./}{figures/}}

\begin{document}

\title{Transmission spectroscopy of the lowest-density gas giant: metals and a potential extended outflow in HAT-P-67b}

\correspondingauthor{Aaron Bello-Arufe}
\email{aaron.bello.arufe@jpl.nasa.gov}

\author[0000-0003-3355-1223]{Aaron Bello-Arufe}
\affil{National Space Institute, Technical University of Denmark, Elektrovej, DK-2800 Kgs. Lyngby, Denmark}
\affil{Division of Geological and Planetary Sciences, California Institute of Technology, 1200 East California Boulevard, Pasadena, CA 91125, USA}
\affil{Jet Propulsion Laboratory, California Institute of Technology, Pasadena, CA 91109, USA}

\author[0000-0002-5375-4725]{Heather A. Knutson}
\affil{Division of Geological and Planetary Sciences, California Institute of Technology, 1200 East California Boulevard, Pasadena, CA 91125, USA}

\author[0000-0002-6907-4476]{João M. Mendonça}
\affil{National Space Institute, Technical University of Denmark, Elektrovej, DK-2800 Kgs. Lyngby, Denmark}

\author[0000-0002-0659-1783]{Michael M. Zhang}
\affil{Department of Astronomy, California Institute of Technology, Pasadena, CA 91125, USA}

\author[0000-0001-9749-6150]{Samuel H. C. Cabot}
\affil{Yale University, 52 Hillhouse Avenue, New Haven, CT 06511, USA}

\author[0000-0002-4227-4953]{Alexander D. Rathcke}
\affil{Center for Astrophysics $|$ Harvard \& Smithsonian, 60 Garden Street, Cambridge, MA 02138, USA}

\author[0000-0001-6424-5005]{Ana Ulla}
\affil{Applied Physics Department, Universidade de Vigo, Campus Lagoas-Marcosende, s/n, E-36310 Vigo, Spain}
\affil{Instituto de Física e Ciencias Aeroespaciais (IFCAE), Universidade de Vigo, Campus de As Lagoas, E-32004 Ourense, Spain}

\author[0000-0003-2527-1475]{Shreyas Vissapragada}
\affil{Division of Geological and Planetary Sciences, California Institute of Technology, 1200 East California Blvd, Pasadena, CA 91125, USA}

\author[0000-0003-1605-5666]{Lars A. Buchhave}
\affil{National Space Institute, Technical University of Denmark, Elektrovej, DK-2800 Kgs. Lyngby, Denmark}



\begin{abstract}
Extremely low-density exoplanets are tantalizing targets for atmospheric characterization because of their promisingly large signals in transmission spectroscopy. We present the first analysis of the atmosphere of the lowest-density gas giant currently known, HAT-P-67b. This inflated Saturn-mass exoplanet sits at the boundary between hot and ultrahot gas giants, where thermal dissociation of molecules begins to dominate atmospheric composition. We observed a transit of HAT-P-67b at high spectral resolution with CARMENES and searched for atomic and molecular species using cross-correlation and likelihood mapping. Furthermore, we explored potential atmospheric escape by targeting H$\alpha$ and the metastable helium line. We detect \ion{Ca}{2} and \ion{Na}{1} with significances of 13.2$\sigma$ and $4.6\sigma$, respectively. Unlike in several ultrahot Jupiters, we do not measure a day-to-night wind. The large line depths of \ion{Ca}{2} suggest that the upper atmosphere may be more ionized than models predict. We detect strong variability in H$\alpha$ and the helium triplet during the observations. These signals suggest the possible presence of an extended planetary outflow that causes an early ingress and late egress. In the averaged transmission spectrum, we measure redshifted absorption at the $\sim 3.8\%$ and $\sim 4.5\%$ level in the H$\alpha$ and \ion{He}{1} triplet lines, respectively. From an isothermal Parker wind model, we derive a mass-loss rate of $\dot{M} \sim 10^{13}~\gs$ and an outflow temperature of $T \sim 9900~\rm{K}$. However, due to the lack of a longer out-of-transit baseline in our data, additional observations are needed to rule out stellar variability as the source of the H$\alpha$ and He signals.

\end{abstract}



\section{Introduction} \label{sec:intro}
Gas giant exoplanets with extremely low densities ($\rho_p \lesssim 0.1~\gcc$) constitute some of the most enticing targets for atmospheric characterization. Their puffy atmospheres are particularly suited for transmission spectroscopy studies, as they filter a larger fraction of the stellar light. Indeed, exoplanet observers have already targeted the transmission spectra of some of these uniquely low-density exoplanets, including WASP-17b \citep[e.g.][]{sedaghati2016,saba2022}, WASP-127b \citep[e.g.][]{chen2018,seidel2020}, KELT-11b \citep[e.g.][]{zak2019,colon2020}, and HAT-P-32b \citep[e.g.][]{damiano2017,alam2020}.

HAT-P-67b \citep{zhou2017} is the lowest-density gas giant currently known. Its mass, $M_p = 0.34~M_{\rm J}$, is similar to that of Saturn, yet its radius is twice as large as Jupiter's, $R_p = 2.085~R_{\rm J}$, which results in a density of only $\rho_p = 0.052~\gcc$. HAT-P-67b is also the largest known transiting exoplanet. Assuming a mean molecular weight of $\mu = 2.3$, HAT-P-67b has a scale height of $H \sim 3500~\rm{km}$ which, together with its relatively bright host star ($V = 10.069$, $J = 9.145$, \citealt{zhou2017}), makes this planet an outstanding target for characterization through transmission spectroscopy.

With an equilibrium temperature of $T_{\rm eq} = 1903~\rm{K}$, HAT-P-67b sits at the transition between hot and ultrahot gas giants. As opposed to their cooler counterparts, ultrahot gas giants exhibit very weak water features in their emission spectra due to a combination of water dissociation and H$^{-}$ opacity \citep{kitzmann2018,lothringer2018,parmentier2018}, and they tend to have thermal inversions in their atmospheres \citep{lothringer2018,pino2020,mansfield2021}. While the atmospheres of hot Jupiters are dominated by the presence of molecular species \citep[e.g.][]{giacobbe2021,carleo2022,guilluy2022}, those of ultrahot Jupiters host a large abundance and diversity of atomic species \citep[e.g.][]{hoeijmakers2019,belloarufe2022tng,kesseli2022,prinoth2022,borsato2023,jiang2023}. 

HAT-P-67b is also an excellent target to search for an escaping atmosphere. This planet is on a close-in 4.8 day orbit around an F-subgiant star and is subject to strong UV irradiation from its host. Also, as pointed out by \citet{zhou2017}, HAT-P-67b has one of the lowest escape velocities among all known exoplanets, $v_{\rm esc}\sim24~\kms$. Escape from HAT-P-67b might potentially be observable in the Balmer H$\alpha$ line of hydrogen, a powerful probe of atmospheric escape \citep[e.g.][]{jensen2012,yan2018,wyttenbach2020}. Alternatively, escape from HAT-P-67b may also be accessible through the 1083~nm triplet of metastable helium \citep[e.g.][]{allart2018,nortmann2018,spake2018}. While \citet{oklopcic2019} argued that planets orbiting hot stars are unlikely to show prominent 1083~nm absorption signals, the recent detection of helium escaping from HAT-P-32b \citep{czesla2022} demonstrates that planets around F-type stars can host observable helium outflows.

In this paper, we present an analysis of the atmosphere of HAT-P-67b using data from CARMENES \citep{quirrenbach2016}: a high-dispersion instrument that simultaneously covers visible and infrared wavelengths. A planet like HAT-P-67b, where atomic and molecular species likely coexist, allows us to exploit the full potential of CARMENES, as we can search for the electronic transitions of atomic species in the visible and for the transitions of molecules at visible and near-infrared wavelengths. Additionally, CARMENES provides simultaneous access to H$\alpha$ and the 1083~nm triplet of metastable helium. Joint modeling of these two signals can improve constraints on the outflow parameters \citep{yan2022}. Lastly, planets receiving high levels of stellar irradiation are often found to host high-velocity winds flowing from their daysides to their nightsides \citep[e.g.][]{snellen2010,casasayasbarris2019,nugroho2020mascara,belloarufe2022tng,kesseli2022,prinoth2022,yzhang2022}, driven by large day-to-night heating variations \citep{showman2013,komacek2016}. Such winds should be resolvable by CARMENES thanks to its high spectral resolution.

This paper is organized as follows. In Section~\ref{sec:observations} we describe our observations and data reduction. Section~\ref{sec:likelihood_mapping} presents our cross-correlation and likelihood analysis to search for atomic and molecular species, and the transmission spectroscopy analysis to search for hydrogen and helium escape. In Section~\ref{sec:results_and_discussion} we report and discuss the results, and Section~\ref{sec:conclusions} includes a summary of the conclusions.

\section{Observations and Data Reduction} \label{sec:observations}
We observed a full transit of HAT-P-67b on the night of 2021 May 17 using CARMENES \citep{quirrenbach2016}, an instrument mounted on the 3.5 m telescope at the Calar Alto Observatory. CARMENES consists of two separate high-resolution echelle spectrographs operating together: the visible (VIS) channel covers wavelengths between 520 and 960~nm with a resolving power of $R=94,600$, and the near-infrared (NIR) channel covers the 960--1710~nm range with $R=80,400$. We set exposure times of 600 and 606 s in the VIS and NIR channels, respectively. These slightly different exposure times account for the difference in readout time between the two detectors, ensuring that exposures are taken simultaneously in both channels. In total, we obtained 37 exposures from each channel: 33 during transit (i.e. those whose midpoint falls between first and fourth contact), two before ingress, and two after egress. The long transit duration of HAT-P-67b (7 hours) limits the amount of out-of-transit baseline at low airmass accessible to ground-based telescopes in any particular night. We started observing HAT-P-67 at an airmass of 1.89, reaching a minimum of 1.01 during transit, and ending the night at 1.12. During the observations, we placed so-called fiber A on the target, and we used fiber B to monitor sky emission.

The data were reduced using version 2.20 of the CARACAL pipeline \citep[CARMENES Reduction And CALibration software,][]{caballero2016}. CARACAL performs dark and bias subtraction, order tracing, flat-relative optimal extraction, and wavelength calibration of both spectral channels, with the wavelengths provided in the rest frame of the observatory and in vacuum. For consistency, we use vacuum wavelengths throughout this work. The signal-to-noise ratios (S/Ns) in the VIS and NIR channels, as measured by CARACAL in the orders that contain H$\alpha$ and the metastable He triplet, were in the range 46--72 and 40--70, respectively.

\section{Data analysis and model comparison}\label{sec:likelihood_mapping}
\subsection{Detrending}
In order to isolate the atmospheric signal of the exoplanet, we first needed to remove the telluric and stellar features from the data. 

\subsubsection{Telluric correction}\label{sec:tellurics}
We used the software tool \texttt{molecfit} to model the telluric absorption lines present in the spectra \citep{smette2015}. \texttt{Molecfit} combines a radiative transfer code with atmospheric profiles to fit the observed spectra, producing a model of the transmission spectrum of the Earth's atmosphere at the time of each observation. Prior to running \texttt{molecfit}, we merged and resampled the different echelle orders onto a common wavelength grid following \citet{belloarufe2022not}. We treated the VIS and NIR channels independently due to their different line spread functions. We then quadratically interpolated the resulting telluric models onto the original order-by-order wavelength grid. We divided each observed spectrum by the corresponding telluric model, effectively correcting the data for telluric absorption. Finally, we inspected all orders visually and masked out the regions where the telluric correction was inadequate or there was excessive noise.

In addition to telluric absorption lines, the spectra from the NIR channel are also markedly contaminated by telluric emission lines. To remove these lines, we first flagged all outliers in the sky spectra from fiber B, and we replaced them by the linear interpolation of their neighboring pixels. We then subtracted the sky spectra from the \texttt{molecfit}-corrected data.

\subsubsection{Correction of the Rossiter-McLaughlin and Center-to-limb Variation Effects}\label{sec:methods_rm_clv_correction}
Inhomogeneities in the spectrum of a star across its disk can significantly impact transmission spectroscopy studies. Of particular relevance to studies at high resolution are the Rossiter-McLaughlin (RM) and center-to-limb variation (CLV) effects. The RM effect occurs when a planet blocks regions of the rotating stellar disk that have different line-of-sight velocities. The CLV effect is the result of changes in the stellar line profiles and intensities between the center and the limb of the stellar disk \citep[][]{yan2017}. The contamination from the RM and CLV effects is exacerbated in systems like HAT-P-67, where the radial velocity of the planet during transit is close to the line-of-sight velocity of the region of the star occulted by the planet \citep[see, e.g.][]{casasayasbarris2022,dethierbourrier2023}.

\begin{figure}
\centering
\includegraphics[width=\linewidth]{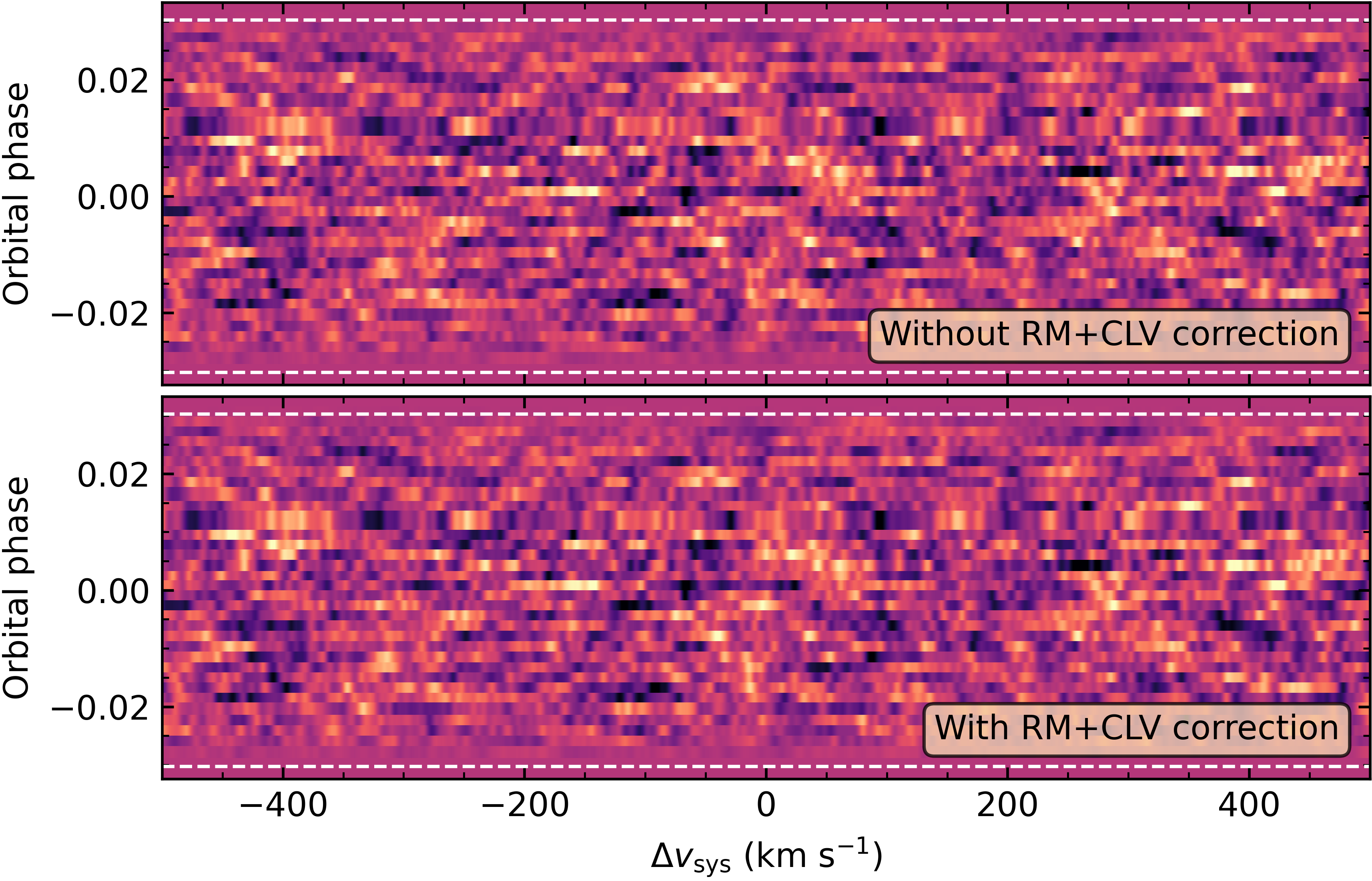}
  \caption{Cross-correlation functions of \ion{Na}{1} without and with the correction of the RM and CLV effects. The horizontal dashed lines indicate the points of first and fourth contact. The RM and CLV contributions are visible in the top plot as a dark slanted trace near the center of the image. This trace is also known as the Doppler shadow, and it is not visible in the lower plot.}
  \label{fig:dopshad_removal}
\end{figure}

We modeled and removed the contribution from the RM and CLV effects to each of our spectra. As in similar works \citep[e.g.][]{yan2017,casasayasbarris2019}, we produced a grid of synthetic stellar spectra at different limb angles using \texttt{Spectroscopy Made Easy} version 522 \citep[][]{piskunov2017}, the Kurucz \texttt{ATLAS9} model atmospheres \citep{castelli2003}, and the Vienna Atomic Line Database \citep{ryabchikova2015}. We discretized the stellar disk into a series of squared cells (in our case, of side length $0.01~R_\star$), and we assigned to each cell a spectrum interpolated from our grid based on the limb angle of such cell. We also Doppler shifted each spectrum according to the line-of-sight velocity of the cell. At the orbital phase of each exposure, we integrated the spectra from the cells blocked by the planetary disk, and divided the resulting spectrum by the full-disk spectrum. We continuum normalized the resulting spectra using a median filter, and we broadened them according to the resolution of the corresponding CARMENES channel. This procedure generates a model of the RM and CLV contributions for each exposure, which we then divided out from the data. In Figure~\ref{fig:dopshad_removal}, we illustrate the effect that the removal of the RM and CLV effects has on the cross-correlation functions (CCFs) of \ion{Na}{1} (see Section~\ref{sec:cc_analysis} for a description of how we calculate the CCFs).

\subsubsection{Removal of the stellar spectra}\label{sec:remove_star_crosscorr}
As a first step to removing the stellar features, we scaled all spectra to the same level of flux. We started by dividing the flux in each pixel by the median of its corresponding exposure and order. We repeated this step one more time after flagging and interpolating over outliers. This iteration ensures that our scaling is not biased by regions with a large number of outliers. 

Next, we aligned all spectra in the rest frame of the HAT-P-67 system. As the spectra were originally in the rest frame of the observatory, this step involved Doppler shifting the wavelength solution of each spectrum by two velocities: the barycentric velocity correction and the velocity of the solar system with respect to the HAT-P-67 system. The barycentric velocity correction accounts for the motion of the observatory around the barycenter of the solar system and is provided by the instrument reduction pipeline in the header of the data files. The velocity of the solar system with respect to the HAT-P-67 system is the negative of the systemic radial velocity, $v_{\rm{sys}}=-1.4~\kms$ \citep{zhou2017}. After shifting all spectra, we interpolated them onto a common wavelength grid. We note that it was not possible to correct for the stellar reflex motion because the radial velocity semiamplitude of the host star is only known to an upper limit of $K_{\star}<36~\ms$ \citep{zhou2017}. However, due to the rotational broadening of the spectra ($\vsini = 35.8\pm1.1~\kms$, \citealt{zhou2017}) and the width of the CARMENES pixels ($\sim 0.9-1.7~\kms$), this limitation should not have a significant impact on the results \citep[e.g.][]{casasayasbarris2019,belloarufe2022tng}.

We removed the stellar features by dividing each spectrum by the average stellar spectrum. During the averaging, each data point was weighted by the inverse of its squared uncertainty (the instrument pipeline provided the uncertainties, which we then propagated through the analysis). Due to the lack of a sufficient number of out-of-transit spectra, we obtained the average stellar spectrum by combining all spectra, including those taken during transit. Including the in-transit spectra in the computation of the average stellar spectrum will only marginally dilute the planetary signal because we have a large number of in-transit spectra, and the CARMENES pixels are narrower than the planet's change in radial velocity between consecutive exposures ($\Delta v \sim 1.7~\kms$).

We used a Gaussian filter with a standard deviation of 200 pixels to remove any variations in the continuum of the spectra. Several factors can cause the continuum to change during the observations, including changes in weather, airmass, and blaze function. After testing different values, we chose a standard deviation of 200 pixels for the Gaussian filter. We then subtracted unity from the spectra such that they were all centered around zero.

\subsubsection{\texttt{SYSREM} as an alternative detrending approach}
As an alternative to \texttt{molecfit}, we also detrended the data from the NIR channel using the \texttt{SYSREM} algorithm \citep{tamuz2005}. Unlike data from the VIS channel, the NIR data are dominated by strong telluric features. The difficulty of accurately matching the shape (including the wings) of these deep lines makes telluric modeling with \texttt{molecfit} more challenging. \texttt{SYSREM} works similarly to principal component analysis, but it can be applied to data with unequal uncertainties. This algorithm was originally introduced to remove systematic effects from stellar light curves and was later adapted to high-resolution spectroscopy studies \citep{birkby2013}, where each pixel is treated as an independent light curve. \texttt{SYSREM} does not rely on any prior knowledge of the systematic effects and, iteratively, it can successfully remove trends that appear linearly in the data, such as stellar and telluric absorption features.

We restricted the use of \texttt{SYSREM} to data from the NIR channel, as \texttt{molecfit} produced a high-quality correction of the telluric lines in the VIS channel. Additionally, while \texttt{SYSREM} effectively removes telluric and stellar features, it can also significantly degrade the planet's signal \citep[e.g.][]{birkby2017,meech2022}.
The number of \texttt{SYSREM} iterations determines the number of trends removed from the data: too many iterations will remove the planetary signal, while too few will leave residual telluric and stellar features. We refer the reader to \citet{cabot2019} and \citet{spring2022} for an in-depth discussion of additional considerations when optimizing \texttt{SYSREM} parameters. As in similar works and in order to avoid bias, we used the same number of \texttt{SYSREM} iterations across all NIR orders \citep[e.g.][]{gibson2020,nugroho2020mascara,spring2022}.

\subsection{Model templates}
Following the methodology presented in \citet{belloarufe2022tng}, we constructed a model of the transmission spectrum for each of the species we were interested in. We assumed an atmosphere of solar metallicity and in chemical equilibrium, and an isothermal profile with a temperature equal to the equilibrium temperature of the planet, $T_{\rm{eq}} = 1903~K$. For each gas, we calculated its concentration with \texttt{FastChem} \citep{stock2018}, and we computed its absorption cross section with \texttt{HELIOS-K} \citep{grimm2021} using the line lists from \citet{kurucz2018} and assuming a Voigt profile for the absorption lines. Our models also include H$^-$ bound-free and free-free absorption \citep{john1988}. At each wavelength, we calculated the transit depth using the formalism in \citet{gaidos2017} and \citet{bower2019}.

We processed the models similarly to the data in order to turn them into templates for the cross-correlation and likelihood analyses. First, we broadened each model according to the instrumental resolution of each CARMENES channel. Then, we removed from each model its continuum, found using a Gaussian filter with a standard deviation equivalent to that of the Gaussian filter applied to the data. 

Our survey focused on atomic and molecular species with a substantial number of detectable lines in the transmission spectrum of a planet like HAT-P-67b and in the CARMENES wavelength range. Our list included \ion{Al}{1}, \ion{Ca}{1}, \ion{Ca}{2}, CO, \ion{Cr}{1}, \ion{Fe}{1}, H$_2$O, \ion{K}{1}, \ion{Li}{1}, \ion{Mg}{1}, \ion{Mn}{1}, \ion{Na}{1}, \ion{Ni}{1}, \ion{O}{1}, OH, \ion{Sc}{1}, \ion{Sc}{2}, \ion{Si}{1}, \ion{Ti}{1}, TiO, \ion{V}{1}, and VO \citep{rothman2010,mckemmish2016,kurucz2018, polyansky2018,mckemmish2019}.

\subsection{Cross-correlation analysis}\label{sec:cc_analysis}
Cross-correlation of high-resolution spectra against model templates has become one of the most successful techniques to robustly identify species in exoplanetary atmospheres. This technique has demonstrated its success in the detection of atomic and molecular species, in both transmission and emission spectroscopy \citep[e.g.][]{snellen2010,brogi2012,hoeijmakers2019,pino2020,kesseli2022}.

For each species, we calculated the corresponding CCFs using the expression in \citet{gibson2020}:
\begin{equation}\label{eq:ccf}
    \textup{CCF} = \sum{\frac{f_i m_i}{\sigma_i^2}},
\end{equation}
where $f_i$ and $m_i$ refer to the values of the spectrum and model template at pixel $i$, and $\sigma_i^2$ is the variance with time of the values in that pixel. The sum spans all pixels in the corresponding CARMENES channel. We Doppler shifted each model template by velocity offsets $\Delta v_{\rm{sys}}$ ranging from $-500$ to 500~\kms\ in steps of 1~\kms. We also scaled the model template by a value between 0 and 1 according to the transit light curve, calculated using \texttt{batman} \citep{kreidberg2015}.

We then generated the cross-correlation $K_p-\Delta v_{\rm{sys}}$ maps. To do so, we Doppler shifted each CCF by a velocity $v_p$:
\begin{equation}\label{eq:vplanet}
    v_p \left(t\right) = K_p \sin{\left(2\pi\phi(t)\right)},
\end{equation}
where $\phi(t)$ is the orbital phase at time $t$, and $K_p$ is the planet's radial velocity semiamplitude, which we allowed to take values between 0 and 500~\kms\ in steps of 1~\kms. For each $K_p$, we summed the Doppler-shifted CCFs and stacked them to generate the $K_p-\Delta v_{\rm{sys}}$ maps. We converted the values in these maps to S/Ns by normalizing them by the standard deviation of those values located at $|\Delta v_{\rm{sys}}| \geq 50~\kms$, far from the exoplanet signal. Any detection should manifest as a peak near $K_{p} \sim 147~\kms$ (the planet's radial velocity semiamplitude according to \citealt{zhou2017}) and $v_{\rm{sys}}=0~\kms$ (as we shifted all spectra to the rest frame of the HAT-P-67 system).

\subsection{Likelihood mapping}\label{sec:cc_to_l}
While classical cross-correlation techniques are highly effective to detect atomic and molecular species, their ability for model comparison is more limited. To overcome this issue, recent works have introduced novel Bayesian retrieval frameworks to extract physical parameters from high-resolution ground-based observations \citep[e.g.][]{brogiline2019,gibson2020}.

In this work, we adopted the statistical framework presented by \citet{gibson2020} to constrain the physical extent of the model atmosphere, in addition to $K_p$ and $\Delta v_{\rm{sys}}$. We computed the log-likelihood as 
\begin{equation}
    \ln{L} = 
    \frac{-N}{2}
    \ln{\frac{\chi^2}{N}
    },
\end{equation}
where $N$ is the number of data points, and $\chi^2$ is given by:
\begin{equation}\label{eq:chisq}
    \chi^2=
    \sum_{i=1}^N{\frac{f^2_i}{\sigma^2_i}} 
    +
    \alpha^2\sum_{i=1}^N{\frac{m^2_i}{\sigma^2_i}}
    -
    2\alpha\sum_{i=1}^N{\frac{f_i m_i}{\sigma^2_i}}.
\end{equation}
Here, $\alpha$ is a parameter that provides information on how well our model captures the physical extent of the atmosphere. This parameter is particularly useful to interpret data from highly irradiated gas giants, as these planets often possess extended atmospheres that models struggle to predict \citep[e.g.][]{hoeijmakers2019}. The final term of this equation is identical to the CCFs from Equation~\ref{eq:ccf}. Therefore, as presented in \citet{nugroho2020mascara}, we calculated the log-likelihood as a function of three parameters ($\alpha$, $K_p$, and $\Delta v_{\rm{sys}}$), which led to a three-dimensional data cube for each species, with these three parameters as the axes. We gave $\alpha$ values from 0.1--10, and we kept the same range of values for $K_p$ and $\Delta v_{\rm{sys}}$ as in the previous section. We subtracted the maximum log-likelihood from each data cube and took the exponential to produce a scaled likelihood, with values ranging from 0--1. 
 

\subsection{Transmission spectroscopy of H$\alpha$ and the metastable He triplet}\label{sec:methods_transmission}

\begin{figure*}
\centering
\includegraphics[width=\linewidth]{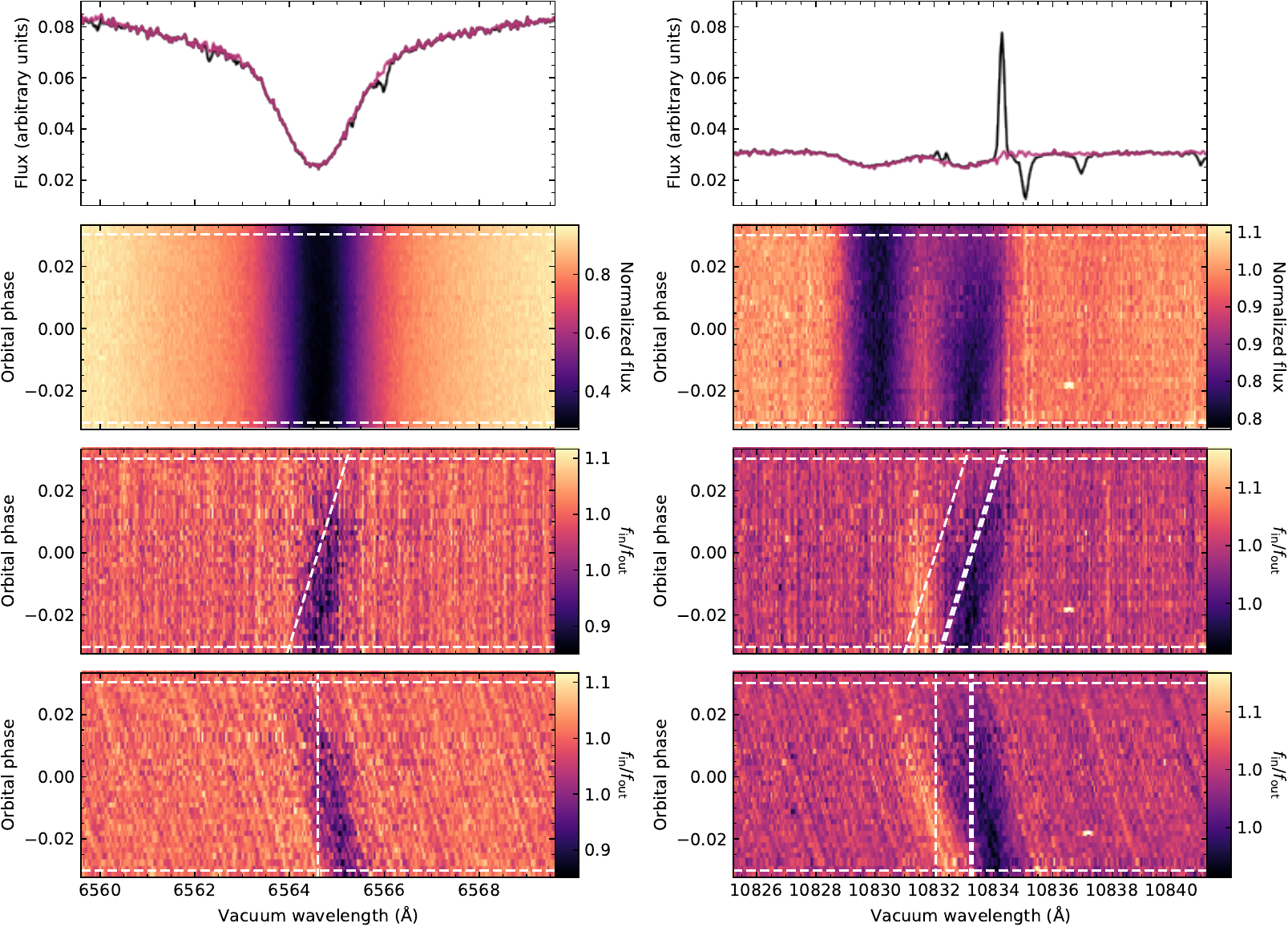}
  \caption{Steps to calculate the transmission spectra in the regions of H$\alpha$ (left panels) and the He triplet (right panels). First row: sample spectrum before (in black) and after (in purple) telluric correction, in the rest frame of the observatory. Second row: telluric-corrected, normalized spectra stacked in the rest frame of the HAT-P-67 system. The horizontal dashed lines indicate the points of first and fourth contact. Third row: same as the second-row panels, but divided by the master stellar spectrum (i.e. the average post-transit spectrum). The slanted lines indicate the theoretical position of the hydrogen and helium lines in the rest frame of the exoplanet. Fourth row: same as the third-row panels, but shifted to the planetary rest frame.}
  \label{fig:ha_he_steps}
\end{figure*}

Here, we describe the steps we followed to search for hydrogen and helium escape in H$\alpha$ and the metastable He triplet. These steps were similar to the ones we followed to prepare the spectra for cross correlation, except for some notable differences in the detrending process.

We removed telluric absorption lines using \texttt{molecfit}, as described in Section~\ref{sec:tellurics}. The spectral regions around H$\alpha$ and the He triplet are mostly affected by water telluric absorption. The removal of these absorption lines for a sample spectrum is shown in the first row of Figure~\ref{fig:ha_he_steps}. We found that \texttt{molecfit} achieved a high-quality modeling of these telluric lines, and hence we did not need to apply the more aggressive \texttt{SYSREM} algorithm.

While the region near H$\alpha$ is free from any significant telluric emission, the sky-monitoring fiber (i.e. fiber B) revealed strong emission lines around the He metastable triplet associated with OH airglow \citep{oliva2015}. In particular, there are two $\Lambda$-split OH doublets from the Q branch that surround the He triplet. The first OH doublet has vacuum wavelengths of 10832.103 and 10832.412~\AA. The second OH doublet, about an order of magnitude stronger than the first one, has wavelengths of 10834.241 and 10834.338~\AA. The two individual components of this second doublet are not resolved by CARMENES and therefore appear as one strong line in the spectra. The top-right panel of Figure~\ref{fig:ha_he_steps} shows the removal of these three telluric emission features from a sample spectrum.

We corrected the OH telluric emission lines using data from fiber B and following a similar modeling approach to \citet{palle2020}. First, we shifted the spectra of the sky-monitoring fiber to a common wavelength grid through a quadratic interpolation, and we combined them into a median sky spectrum. We then fit three Gaussian functions to the three resolved OH lines, keeping the amplitudes and widths of the two components of the resolved doublet equal to each other. We scaled this model to match the amplitude of the strongest OH feature in each individual sky spectrum, allowing for a small shift in the position of the peaks. Because of the difference in throughput between both fibers, we multiplied the individual sky models by a scaling factor of 0.881 prior to subtracting them from the science spectra. We found that this approach to remove the OH telluric lines provided slightly cleaner spectra in this region than just directly subtracting the sky spectra as described in Section~\ref{sec:tellurics}.

We then removed a model of the RM and CLV contributions following Section~\ref{sec:methods_rm_clv_correction} and normalized the resulting spectra. As in \citet{palle2020}, we normalized the data by dividing each individual spectrum in the order that contained the He triplet by the median value in a region of the same order but near the continuum, far from any stellar lines. We repeated this step in the order that contained H$\alpha$. We shifted the normalized spectra from the rest frame of the observatory to that of the HAT-P-67 system, as we did in the cross-correlation analysis (Section~\ref{sec:remove_star_crosscorr}). After shifting all spectra, we interpolated them onto a common wavelength grid. The second row of Figure~\ref{fig:ha_he_steps} shows the normalized spectra in the rest frame of the HAT-P-67 system.

We computed the transmission spectra by dividing the spectra by the master stellar spectrum. As we might be probing the escaping atmosphere, any potential H and He signal may not necessarily follow the planetary orbital velocity and might be broader in wavelength than those originating from the bounded atmosphere. For that reason, we did not include any in-transit spectra in this calculation of the master stellar spectrum. Additionally, as will be evidenced by the transmission light curves later in the analysis, there may be planetary absorption prior to the point of first contact. Therefore, we calculated the master stellar spectrum by averaging only the spectra taken after transit. We show the resulting transmission spectra in the third row of Figure~\ref{fig:ha_he_steps}.

Finally, we obtained the master transmission spectrum of HAT-P-67b by taking the weighted average of all the transmission spectra during transit in the planetary rest frame. The last row of Figure~\ref{fig:ha_he_steps} shows the transmission spectra stacked in the planetary rest frame.

\section{Results and Discussion}\label{sec:results_and_discussion}
\subsection{Detection of Ca II and Na I}
\begin{figure*}
\centering
\includegraphics[width=\linewidth]{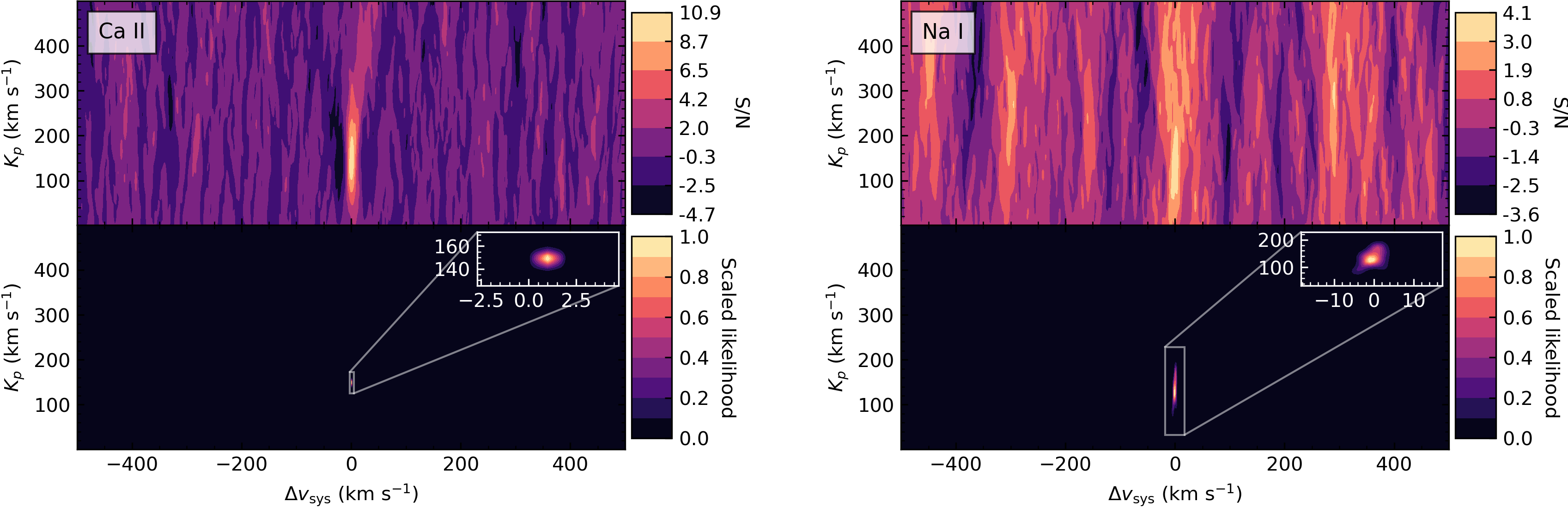}
  \caption{Cross-correlation and likelihood maps of the species we detected in the atmosphere of HAT-P-67b. For each species, we present the cross-correlation map (above) and likelihood map (below). The likelihood maps correspond to the value of $\alpha$ in our data cube that maximizes the scaled likelihood. In the likelihood maps, we also include an inset that zooms in on the region where the scaled likelihood peaks. Note that the insets do not have the same extent.}
  \label{fig:species_detections}
\end{figure*}

The cross-correlation analysis reveals the presence of \ion{Ca}{2} and \ion{Na}{1} in the atmosphere of HAT-P-67b. Figure~\ref{fig:species_detections}
presents the cross-correlation $K_p-\Delta v_{\rm{sys}}$ maps for each of these species. The S/N of these detections, as measured at the peak of the signals, is 10.9 in the case of \ion{Ca}{2}, and 4.1 in the case of \ion{Na}{1} (see Table~\ref{tab:likelihood}).

\begin{table*}
\centering
\caption{Retrieved Values of the Planet's Radial Velocity Semiamplitude ($K_p$), the Radial Velocity Offset ($\Delta v_{\rm sys}$), and the Scale Factor ($\alpha$) from the Likelihood Analysis for the Two Metals we Detected in the Atmosphere of HAT-P-67b.}\label{tab:likelihood}
\begin{tabular}{l|ccccc}
\hhline{======}
Species  & $K_p$ (\kms) & $\Delta v_{\rm sys}$ (\kms) & $\alpha$ & S/N & Sig. ($\sigma$) \\
\hline
\ion{Ca}{2} & $148.8 \pm 4.7$ & $1.02 \pm 0.3$ & $5.97 \pm 0.45$ & 10.9 & 13.2 \\
\ion{Na}{1} & $130 \pm 20$ & $-0.5 \pm 1.8$ & $0.49 \pm 0.11$ & 4.1 & 4.6 \\
\hline
\end{tabular}
\tablecomments{For each species, we also include the S/N from the cross-correlation $K_p-\Delta v_{\rm sys}$ maps, as well as the significance of the detection as calculated by dividing $\alpha$ by its uncertainty.}
\end{table*}

Switching from cross-correlation to likelihood maps, which are also shown in Figure~\ref{fig:species_detections}, allows us to place constraints on $K_p$, $\Delta v_{\rm{sys}}$ and $\alpha$ for each species. We present these results in Table~\ref{tab:likelihood}. As in \citet{nugroho2020mascara}, we calculated these parameters by fitting a Gaussian function to the slice that goes through the maximum scaled likelihood of the data cubes described in Section~\ref{sec:cc_to_l}. The standard deviation of the Gaussian function gives us an estimate of the uncertainty of the parameter.

The retrieved $K_p$ and $\Delta v_{\rm{sys}}$ values of the \ion{Ca}{2} and \ion{Na}{1} detections are consistent with the expected orbital values. The planetary radial velocity semiamplitude ($K_p \approx 147~\kms$, \citealt{zhou2017}) is within $1\sigma$ of the retrieved $K_p$ values for both detections. Unlike in many other hot and ultrahot gas giants \citep[e.g.][]{snellen2010,casasayasbarris2019,nugroho2020mascara,belloarufe2022tng,kesseli2022,prinoth2022,yzhang2022}, we do not measure the blueshifted absorption often associated with day-to-night winds.

The values we retrieve for $\alpha$ inform us of the extent of the atmosphere associated with each species. As shown in Table~\ref{tab:likelihood}, we find that the retrieved \ion{Ca}{2} signal has the largest value of $\alpha$; our model underestimates the line contrast of this species by a factor of $\sim 6$. This is similar to results in high-resolution studies of other highly irradiated planets \citep[e.g.][]{yan2019,nugroho2020mascara,borsa2021,tabernero2021,belloarufe2022tng}, where ionized calcium is often found at very high altitudes, potentially due to photoionization or the presence of an escaping envelope.

Meanwhile, our models overestimate the line contrast of \ion{Na}{1} by a factor of 2. In our analysis, we broaden the model templates according to the instrumental resolution, but we ignore other line-broadening mechanisms, such as orbital smearing and rotational broadening, which can bias the retrieved value of $\alpha$. In each exposure, the planet changes radial velocity by only $\sim 1~\kms$, so we do not expect orbital smearing to significantly alter $\alpha$. However, planetary rotation can have a measurable effect on the line depths. For example, rotational broadening of WASP-76b can reduce its line depths by a factor of $\sim 2$ \citep{casasayasbarris2021}. While rotational broadening in HAT-P-67b ($v_{\textup{rot}}\sim 2~\kms$, assuming tidal locking) is less severe than in WASP-76b ($v_{\textup{rot}}\sim 5~\kms$), this mechanism may partially explain why the \ion{Na}{1} lines of HAT-P-67b are shallower than we expected. Other factors can further reduce line contrasts. Silicate clouds, which our models ignore, might start to form in planets with an equilibrium temperature similar to that of HAT-P-67b \citep[e.g.][]{lothringer2022}. These clouds would lead to muted spectral features. Atmospheric temperature and metallicity can also influence the depths of the spectral lines, as they affect the extent of the atmosphere and the optical depth, respectively.

\begin{figure*}
\centering
\includegraphics[width=\linewidth]{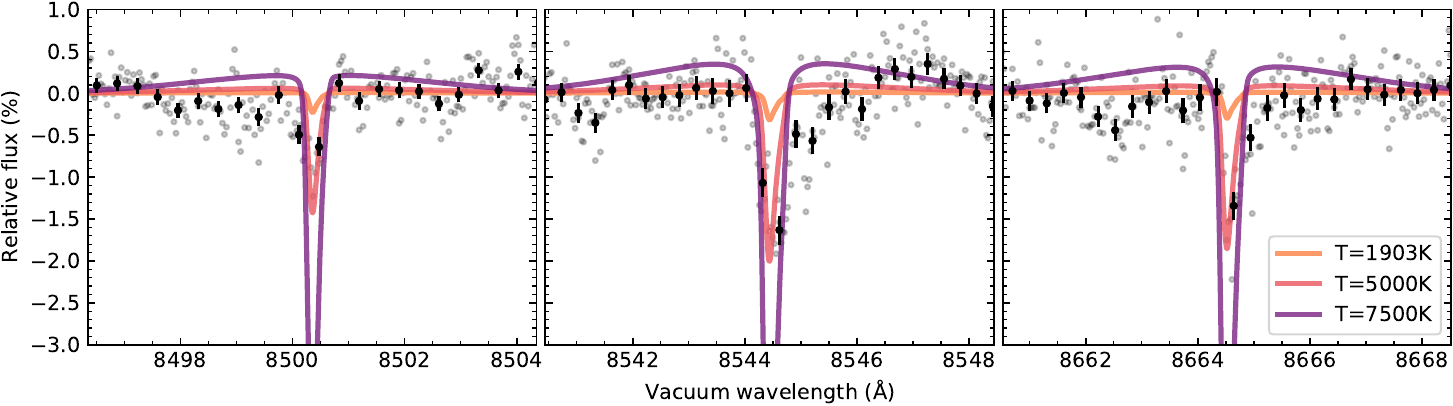}
  \caption{The transmission spectrum of HAT-P-67b in the \ion{Ca}{2} infrared triplet (IRT) lines and the effect of different degrees of ionization on the modeled line depths. The gray and black data points show the unbinned and binned spectra, respectively, average-combined in the planetary rest frame. The different colored lines show models computed at different atmospheric temperatures to simulate various levels of ionization.}
  \label{fig:ca_irt_comparison}
\end{figure*}

Nonetheless, the fact that we only underestimate the line depths of the ionic species (\ion{Ca}{2}) suggests that the lower-density atmosphere of HAT-P-67b may be more ionized than the models predict. In Figure~\ref{fig:ca_irt_comparison}, we study the effect of a higher degree of ionization on the transmission spectrum of HAT-P-67b.

Figure~\ref{fig:ca_irt_comparison} shows the transmission spectrum near the \ion{Ca}{2} infrared triplet (IRT), obtained following the steps in Section~\ref{sec:methods_transmission}. Overlaid are model templates calculated at various temperatures with the goal of simulating the effects of different ionization levels. The model template used in our analysis, calculated at the equilibrium temperature of the planet (1903~K), underestimates the \ion{Ca}{2} line depths by several-fold, in agreement with the $\alpha$ value we found for this species. However, hotter (and therefore more ionized) models can reproduce the observed \ion{Ca}{2} IRT line depths, demonstrating that the upper atmosphere of HAT-P-67b may be more ionized than we predict.

Dividing the values of $\alpha$ by their uncertainties provides us with an estimate of the detection significance of each species \citep[e.g.][]{gibson2020,nugroho2020mascara}. The results are shown in Table~\ref{tab:likelihood}. We find significances of $13.2~\sigma$ (\ion{Ca}{2}) and $4.6~\sigma$ (\ion{Na}{1}). Different metrics are used in high-resolution spectroscopy studies to estimate the significance of a detection, including S/N from cross-correlation maps \citep[e.g.][]{brogi2012}, Welch's T-tests that compare the distribution of cross-correlation values in and out of the planet's trail \citep[e.g.][]{brogi2012,birkby2013}, and, more recently, likelihood maps \citep[e.g.][]{brogiline2019,gibson2020}. Works like \citet{cabot2019} and \citet{spring2022} have discussed the robustness of some of these metrics. Here, we find that the significances derived from our likelihood analysis are consistent with the S/N estimates from the cross-correlation maps.

\begin{figure*}
\centering
\includegraphics[width=\linewidth]{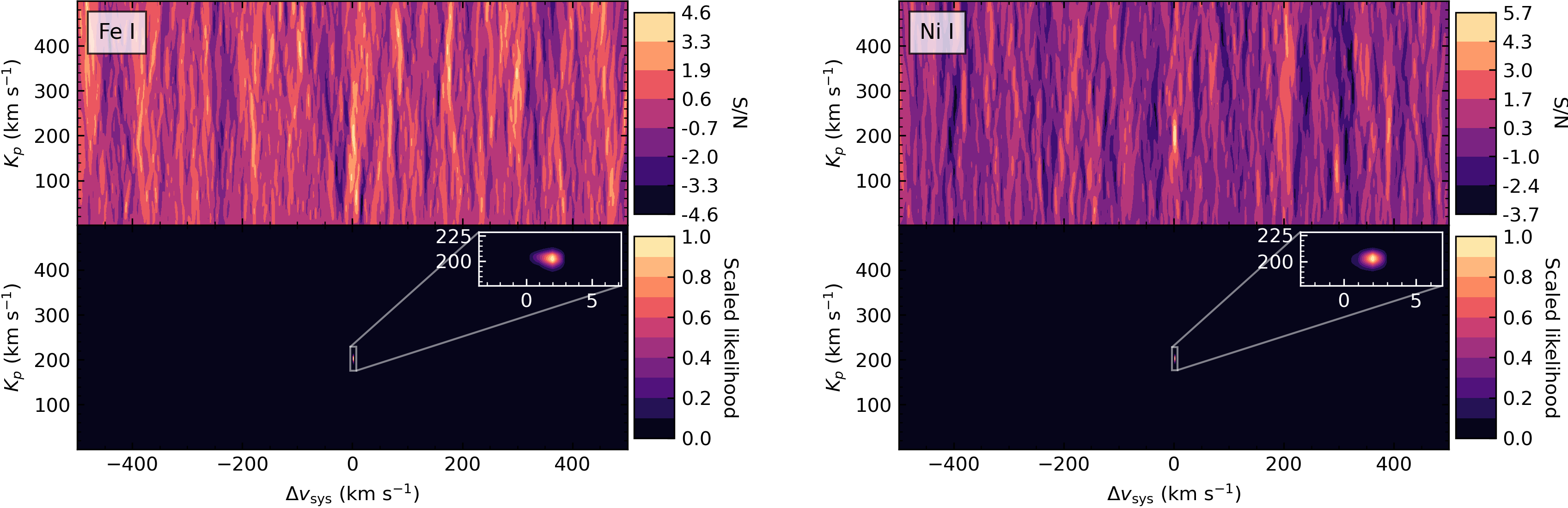}
  \caption{Same as Figure~\ref{fig:species_detections}, but for the species that are likely artifacts of the RM and CLV correction and not true atmospheric detections.}
  \label{fig:species_detections_tentative}
\end{figure*}

We also found a strong peak near the expected velocity of the planet in the $K_p-\Delta v_{\rm{sys}}$ maps of two additional species, \ion{Fe}{1} and \ion{Ni}{1}, as shown in Figure~\ref{fig:species_detections_tentative}.
However, the location of these peaks indicates that they are likely artifacts of the RM and CLV correction rather than true atmospheric signals. HAT-P-67b has a radial velocity semiamplitude of $K_{p} \sim 147~\kms$, but the \ion{Fe}{1} and \ion{Ni}{1} signals peak at $K_{p} \approx 202\pm 5~\kms$ (see Table~\ref{tab:likelihood_tentative}), which coincides with the velocities at which we inject the modeled RM and CLV spectra in Section~\ref{sec:methods_rm_clv_correction}. Additionally, we re-ran our analysis without implementing the RM and CLV correction, and we found that the \ion{Fe}{1} and \ion{Ni}{1} peaks vanished. Therefore, we conclude that the \ion{Fe}{1} and \ion{Ni}{1} peaks are spurious signals caused by an overcorrection of the RM and CLV effects.

\begin{table*}
\centering
\caption{Same as Table~\ref{tab:likelihood}, but for the Species That Are Likely Artifacts of the RM and CLV Correction and Not True Atmospheric Detections.}\label{tab:likelihood_tentative}
\begin{tabular}{l|ccccc}
\hhline{======}
Species  & $K_p$ (\kms) & $\Delta v_{\rm sys}$ (\kms) & $\alpha$ & S/N & Significance ($\sigma$) \\
\hline
\ion{Fe}{1} & $202.3 \pm 5.3$ & $1.80 \pm 0.54$ & $0.354 \pm 0.081$ & 4.6 & 4.4 \\
\ion{Ni}{1} & $202.5 \pm 5.1$ & $1.90 \pm 0.49$ & $1.05 \pm 0.19$ & 5.7 & 5.5 \\
\hline
\end{tabular}
\end{table*}

\subsection{Hydrogen and helium escaping from HAT-P-67b?}
\begin{figure*}
\centering
\includegraphics[width=\linewidth]{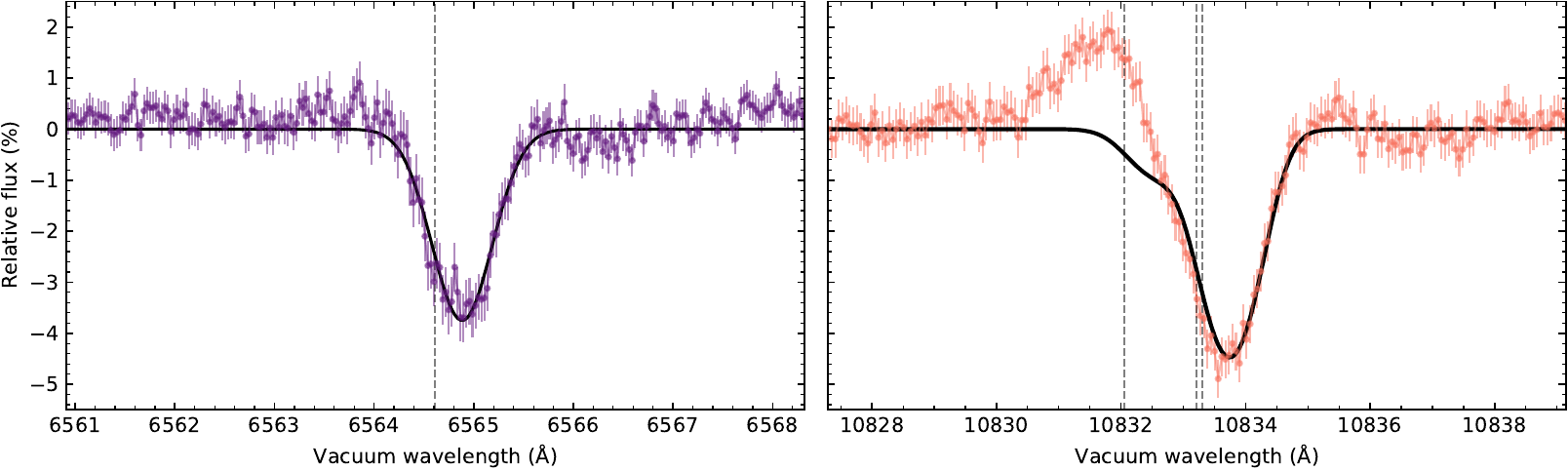}
  \caption{Combined transmission spectrum in the planet's rest frame near H$\alpha$ (left) and the He triplet (right). The vertical dashed lines mark the wavelengths of the H$\alpha$ and He triplet lines. For H$\alpha$ we also plot a Gaussian fit. For the He triplet, we include the transmission spectrum produced by the retrieved Parker wind model.}
  \label{fig:master_in}
\end{figure*}

Visual inspection of the transmission spectra in the last row of Figure~\ref{fig:ha_he_steps} reveals strong absorption near H$\alpha$, as well as strong emission and absorption near the He triplet. While the H$\alpha$ signal is consistent with what we might expect from an extended hydrogen envelope around HAT-P-67b, the emission feature in the transmission spectra near the He triplet could be a manifestation of stellar variability \citep[e.g.][]{palle2020aumic,feinstein2021}. Figure~\ref{fig:master_in} shows the average, in the planet's rest frame, of all in-transit transmission spectra in the regions near H$\alpha$ and the He triplet. We measure an average flux deficit at the $\sim 3.8\%$ and $\sim 4.5\%$ level, respectively, redshifted from the predicted line position in both cases, while the blue wing of the He feature shows an emission bump at the $\sim$2\% level. If the absorption is planetary, these line depths would correspond to an equivalent height of the atmosphere for each species of $\delta R_{p,\,\rm{H\,I}} \sim 1.5R_{p}$ and $\delta R_{p,\,\rm{He\,I}} \sim 1.7R_{p}$. Normalizing these values by the scale height of HAT-P-67b, we obtain $\delta R_{p,\,\rm{H\,I}}/H \sim 65$ and $\delta R_{p,\,\rm{He\,I}}/H \sim 73$, comparable to what \citet{czesla2022} found for HAT-P-32b.

\begin{figure}
\centering
\includegraphics[width=\linewidth]{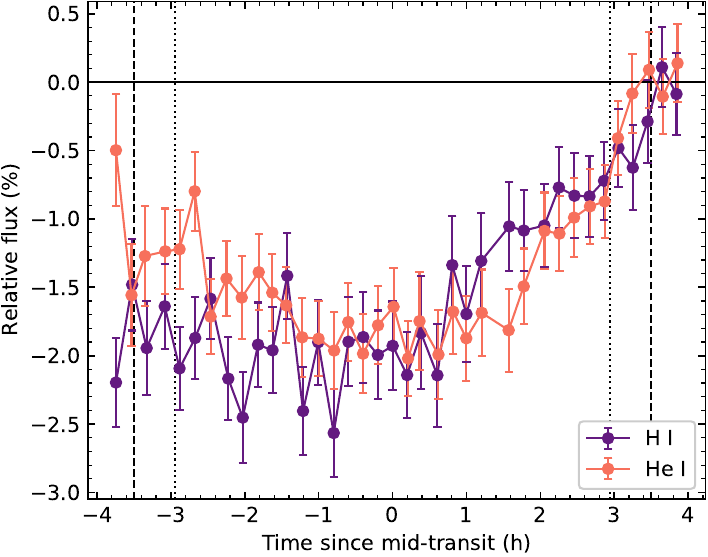}
  \caption{Transmission light curves of H$\alpha$ and the He triplet in bandpasses of widths 2~\AA\ and 3.3~\AA, respectively. The dashed lines indicate the points of first and fourth contact of the white light transit, while the dotted lines mark the points of second and third contact. The horizontal line indicates the baseline, i.e. no absorption.}
  \label{fig:lightcurves}
\end{figure}

We calculated transmission light curves to study the time dependence of the hydrogen and helium signals. We took an average of the relative flux of the transmission spectra within narrow passbands centered on H$\alpha$ (6564.61~\AA) and the He triplet (10833.22~\AA) in the planet rest frame. During the averaging, we weighted each pixel by its width. The bandpasses of H$\alpha$ and the He triplet had widths of 2 and 3.3~\AA, respectively, such that they corresponded to equal widths in velocity space. Figure~\ref{fig:lightcurves} presents the light curves of the H$\alpha$ and He triplet signals. These light curves show that there is strong absorption before the start of the optical transit, and they justify our choice of only using the post-transit spectra as baseline to remove the stellar features (see Section~\ref{sec:methods_transmission}).

We offer three possible scenarios to explain the He signal seen in the transmission spectrum of HAT-P-67b:

\textit{Scenario 1: The signal is of planetary origin, and all ``out of transit" spectra are contaminated by planetary absorption.} As described in Section~\ref{sec:observations}, only four out of the 37 spectra in our data set were taken outside the optical transit: two before the point of first contact, and two after the point of fourth contact. In total, we have $\sim20$ minutes of pre-transit observations, and $\sim20$ minutes of post-transit data. The transmission light curves show strong absorption in the pre-transit spectra. However, the presence of planetary absorption also in the post-transit spectra (i.e. the baseline we used to remove the stellar signal) could result in a spurious emission signal. A helium outflow with a radius of $R_{\rm env} \sim 2.2 R_p \sim 4.5 R_{\rm J}$ would lead to a transit in the He 10833~\AA\ triplet that starts 20 minutes earlier and finishes 20 minutes later.
    
Planetary helium absorption outside the optical transit due to an outflowing envelope has been reported in other planets. For example, \citet{nortmann2018} found post-transit absorption in the Saturn-mass planet WASP-69b that extends $\sim 22$ minutes beyond the end of the optical transit, which they interpret as helium escape from a comet-like envelope that trails behind the planet. Similarly, \citet{spake2021} and \citet{zhang2023} detected significant post-egress absorption in the warm gas giant WASP-107b and the mini-Neptune TOI-2076b, respectively. Meanwhile, \citet{czesla2022} reported redshifted absorption before the optical transit of HAT-P-32b---a gas giant that, like HAT-P-67b, orbits an F-type star on a close-in orbit. \citet{zhang2023} also reported pre-transit absorption in the young mini-Neptune TOI 1430.01, although they do not rule out stellar variability as the source of the signal. An up-orbit stream \citep[e.g.][]{lai2010,carrollnellenback2017,mccann2019} can plausibly explain pre-transit absorption. In an up-orbit stream, material flows toward the star through a nozzle at the L2 point, which leads to (redshifted) pre-ingress absorption that is stronger than (blueshifted) post-egress absorption, a behavior consistent with what we see in our data. \citet{macleodoklopcic2022} found that absorption both before and after the optical transit may be explained by moderate levels of impingement of the stellar wind on the planetary outflow. They also pointed out that, in this scenario of moderate stellar wind interaction, there is a component of absorption whose velocity is constant in the stellar rest frame. Therefore, the He signal not tracing the planetary motion does not necessarily rule out the planet as the source of the absorption.
    
\begin{figure}
\centering
\includegraphics[width=\linewidth]{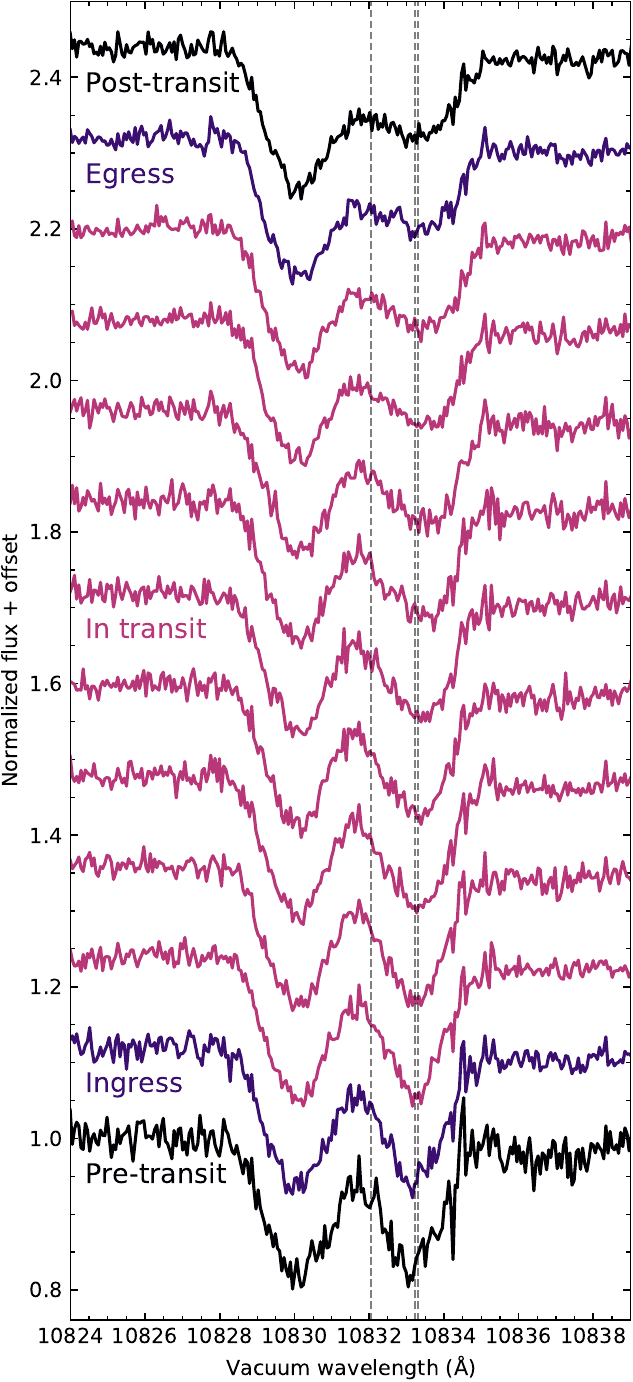}
  \caption{Evolution of the stellar spectra in the metastable He triplet region during the observations. The first spectrum from the bottom is the median pre-transit spectrum, while the first spectrum from the top is the median post-transit spectrum. In the middle, with time increasing from bottom to top, we show the spectra during the transit, median-combined in groups of three for visualization purposes. The vertical dashed lines mark the location of the three components of the He triplet.}
  \label{fig:evolution}
\end{figure}

\textit{Scenario 2: The signal is due to stellar variability.} Without a longer out-of-transit baseline, we cannot rule out stellar variability as the source of the changes seen in the He line. In Figure~\ref{fig:evolution} we plotted the evolution of the stellar spectrum in the region of the Si line (near 10830~\AA) and the He triplet during the observations, with time going upwards. This is similar to the second panel of Figure~\ref{fig:ha_he_steps}, but it allows us to better visualize any changes in the stellar line profiles. Additionally, in Figure~\ref{fig:evolution} we median-combined the host star spectra in trios (in pairs in the case of the out-of-transit spectra) to decrease the noise in the plot. The Figure shows how the median of the two pre-transit spectra (lowest line) differs greatly from the median of the two post-transit spectra (highest line). Before the transit, the stellar Si line and the He triplet form a ``W" shape, while after the transit, the He triplet is significantly deformed. This means that if the signal is of stellar origin, it cannot be due to the contrast effect \citep[e.g.][]{cauley2018}, whereby a planet transits a heterogeneous stellar disk, with starspots and faculae, leading to spurious features in the transmission spectrum. Instead, the signal must be caused by changes in the disk-integrated stellar spectrum. The star is a fast rotator, so these changes could in principle be induced by the rotation of the stellar surface over the course of the observations.

The change of the stellar He triplet profile outside the optical transit also rules out that the He signal we observe in the transmission spectra is due to uncorrected RM and/or CLV effects. Additionally, the contribution from these effects is well below the level of variability we measure in the data. We note that the methodology we applied to correct for these effects is not applicable to the He triplet, as photospheric models do not include chromospheric lines, and the limb-angle dependence of this line is unknown \citep[e.g.][]{czesla2022}. However, the contamination at the He triplet should be lower than at H$\alpha$ due to the shallower stellar line.

Figure~\ref{fig:evolution} also shows how it is not only the core of the He line that changes during the observations but also the wings. In the stellar rest frame, the variability of the stellar spectra spans $\sim 3.25$~\AA, which at these wavelengths corresponds to $\sim 90~\kms$. This value is larger than the change in radial velocity of the planet during the observations ($\sim 60~\kms$) and larger than the FWHM of observed planetary helium outflows \citep[$\sim 30~\kms$; e.g.][]{zhang2023}.
    
The \ion{Ca}{2} infrared triplet is a good diagnostic of chromospheric activity \citep[e.g.][]{linsky1979}. In our data, the \ion{Ca}{2} IRT lines are dominated by planetary absorption rather than showing indications of stellar activity. We know that the absorption is planetary because the \ion{Ca}{2} cross-correlation map, mainly driven by the deep \ion{Ca}{2} IRT lines in the model template, has a strong peak at precisely the radial velocity semiamplitude of the planet (Figure~\ref{fig:species_detections}). In addition, we do not see any other significant features in the \ion{Ca}{2} cross-correlation map that might be caused by stellar variability.

\textit{Scenario 3: The signal is due to a combination of planetary absorption and stellar variability.} The two previous scenarios are not mutually exclusive, and the He signal we see in the data could result from the combination of both.

Given that we are not able to confidently rule out stellar variability as the source of the He signal, we must also question the nature of the H$\alpha$ signal. \citet{guilluy2020} found an anticorrelation between the night-to-night variations of the out-of-transit fluxes of HD\,189733 at the core of the H$\alpha$ and He lines. While we find the opposite effect in HAT-P-67 (i.e. the depths of both stellar lines decrease during the observations), we do not rule out stellar variability as the source of the H$\alpha$ signal, as this line is also sensitive to changes in the chromosphere. However, we can confidently claim that the source of both H$\alpha$ and He signals must be astrophysical. The width and amplitude of the variability and the lack of changes in neighboring lines rule out data reduction artifacts such as the correction of telluric lines, cosmic rays, and bad pixels, or the continuum normalization.

\begin{figure}
\centering
\includegraphics[width=\linewidth]{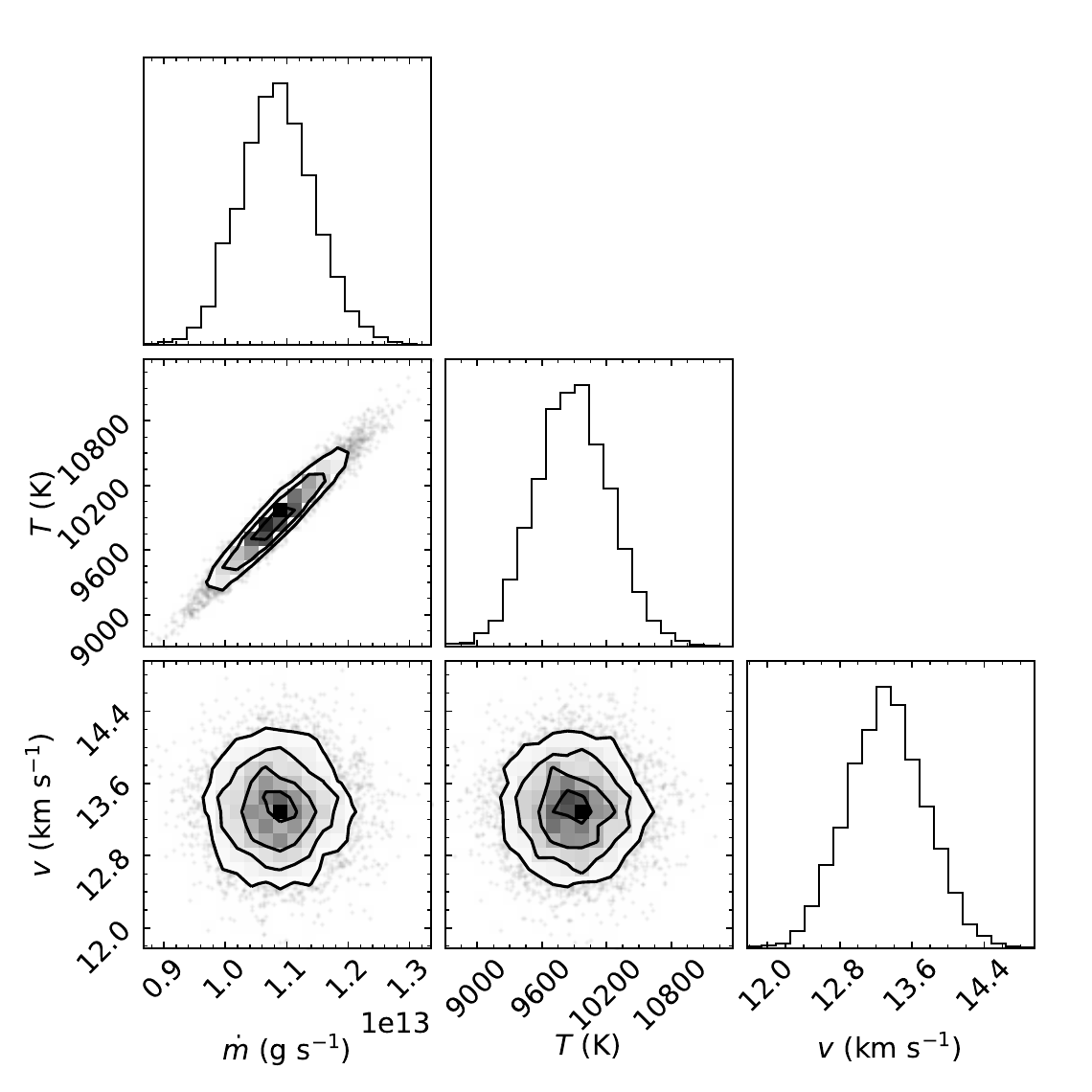}
  \caption{Posterior distributions of the mass-loss rate, outflow temperature, and outflow line-of-sight bulk velocity assuming that the He signal in the transmission spectrum of HAT-P-67b is planetary in nature. }
  \label{fig:posteriors_p-winds}
\end{figure}

If we assume that the He signal is caused by a planetary outflow, we can fit the transmission spectrum in Figure~\ref{fig:master_in} with a one-dimensional, isothermal, H+He Parker wind model to constrain the mass-loss rate, the outflow temperature, and its line-of-sight bulk velocity. To this end, we use \texttt{p-winds} \citep{dossantos2022}, a Python implementation of the model described by \citet{oklopcichirata2018} and \citet{lampon2020}. For our model, we assume no limb darkening and a 90\% H and 10\% He atmosphere. There are no available measurements of the X-ray or EUV flux of HAT-P-67, so we use the monochromatic fluxes relative to the He and H ionization edges of HAT-P-32 \citep{czesla2022}, a star whose effective temperature (\teff) closely resembles that of HAT-P-67. For the monochromatic flux relative to the He triplet ionization edge (911--2593~\AA), we integrate the flux of a photospheric model \citep{castelli2003} with the same \teff, metallicity (Fe/H) and surface gravity (\logg) as HAT-P-67. In the fits, we used the spectra in a 10~\AA\ region centered on the He feature, without masking the bump near 10832~\AA. We retrieve a mass-loss rate of $\dot{M}=1.083^{+0.061}_{-0.058}\times 10^{13}~\gs$, an outflow temperature of $T=9890^{+350}_{-340}~\rm{K}$, and an outflow line-of-sight bulk velocity of $v=13.32 \pm 0.43~\kms$. The posterior distributions are shown in Figure~\ref{fig:posteriors_p-winds}, and the retrieved model transmission spectrum in Figure~\ref{fig:master_in}.

Assuming that the only source of heat powering the outflow is photoionization of hydrogen and helium, we can estimate the maximum mass-loss rate for our retrieved isothermal Parker wind model. Based on the calculations presented in \citet{vissapragada2022} and implemented in \texttt{p-winds}, we derive a maximum mass-loss rate an order of magnitude larger than the retrieved mass-loss rate. This means that our Parker wind model is energetically self-consistent, and photoionization alone can provide enough energy to power it.

In Figure~\ref{fig:master_in} we also include a Gaussian fit to the H$\alpha$ signal to compare its velocity shift to that of the modeled helium outflow. From the Gaussian toy model, we obtain a Doppler shift of $v=12.36 \pm 0.56~\kms$, similar to the offset measured in the He signal ($v=13.32 \pm 0.43~\kms$). We must remember though that if the signals are of planetary origin, we are likely using a contaminated out-of-transit baseline. This means that future observations with uncontaminated out-of-transit baselines may find a different line profile, meaning that the retrieved mass loss rate, outflow temperature, and outflow velocity will change.

\section{Conclusion}\label{sec:conclusions}
We presented the first analysis of the atmosphere of the extremely low-density Saturn-mass planet HAT-P-67b. Using high-resolution spectra obtained with the CARMENES instrument of a single transit event, we report the following findings:

\begin{enumerate}
    \item We detect neutral sodium and ionized calcium in the transmission spectrum of HAT-P-67b, with significances of $4.6\sigma$ and $13.2\sigma$, respectively. Its equilibrium temperature places HAT-P-67b at the transition between hot and ultrahot Jupiters, and the detection of \ion{Ca}{2} in its atmosphere is particularly interesting, as it is generally found in ultrahot Jupiters. To our knowledge, HAT-P-67b is the coldest exoplanet so far with a detection of \ion{Ca}{2}.
    
    \item The likelihood-mapping analysis reveals that our hydrostatic, isothermal, and chemical-equilibrium models largely underpredict the line contrast of \ion{Ca}{2}. We demonstrate that this may be due to the lower-density atmosphere of HAT-P-67b being more ionized than our models predict.
    
    \item We find signals from two additional species, \ion{Fe}{1} and \ion{Ni}{1}, near the orbital velocity of the planet. However, these are likely spurious signals introduced during the removal of the RM and CLV effects. Particularly in the case of planets whose orbital motion partially overlaps with their Doppler shadow, careful inspection of the $K_p-\Delta v_{\rm sys}$ maps may help prevent false positives.
    
    \item We observe strong variability in the H$\alpha$ and metastable He triplet lines. Unfortunately, due to the lack of a significant out-of-transit baseline, we were not able to confirm if these changes are due to an H/He planetary outflow or to stellar variability.
\end{enumerate}

Ultimately, the only way to confirm that the H$\alpha$ and metastable He triplet signals are of planetary origin is with a second transit observation. On the night of 2022 March 11, we attempted to reobserve the transit of HAT-P-67b with CARMENES, this time with a significantly longer pre-transit baseline, but unfortunately bad weather conditions prevented this observation from happening. If a second observation replicated our original strategy (i.e. full transit with minimal data outside optical transit) and found the same behavior in the H$\alpha$ and He lines, that would constitute convincing evidence that the signal is planetary in nature. However, accurately measuring the total amplitude and duration of the absorption signal would require a longer out-of-transit baseline. Due to the long transit duration, combining partial transits from different nights (or from different sites on the same night) may be necessary. The variability we find in H$\alpha$ and the metastable He triplet are so strong that it should also be accessible to relatively small facilities, such as the 2.56\,m Nordic Optical Telescope in the case of H$\alpha$ \citep{belloarufe2022not}. If the presence of an outflow is confirmed, characterization of the high-energy spectrum of the host star with space telescopes would greatly inform modeling efforts.

With its extremely low density and a uniquely large-scale height, HAT-P-67b is an exceptional target to characterize the transition region between hot and ultrahot giants. Additionally, if its outflow is confirmed, HAT-P-67b may become a prime target for simultaneous characterization of two probes of atmospheric escape \citep[e.g.][]{czesla2022,yan2022} and for studies of the morphology of hydrodynamic winds \citep[e.g.][]{mccann2019,macleodoklopcic2022}.


\vspace{15mm}
We thank the anonymous referee for providing a thoughtful report that improved the quality of this work. We are also grateful to Lisa Nortmann and the rest of the CARMENES Atmospheres Legacy Group, who kindly and swiftly offered their telescope time in the hope of observing a second transit of HAT-P-67b. We also thank the staff at Calar Alto Observatory, including Ignacio Vico for carrying out the observations, and Gilles Bergond for valuable help during our program. A.B.-A. gratefully acknowledges support from the Niels Bohr Foundation under the Royal Danish Academy of Sciences and Letters. A.U. gratefully acknowledges grant PR2015-00511 from the Spanish Ministry of Education, Culture, and Sports. This work is based on observations collected at the Centro Astronómico Hispano en Andalucía (CAHA) at Calar Alto, operated jointly by Junta de Andalucía and Consejo Superior de Investigaciones Científicas (IAA-CSIC). Part of this research was carried out at the Jet Propulsion Laboratory, California Institute of Technology, under a contract with the National Aeronautics and Space Administration (80NM0018D0004). This research has made use of the NASA Exoplanet Archive, which is operated by the California Institute of Technology, under contract with the National Aeronautics and Space Administration under the Exoplanet Exploration Program.

%

\vspace{5mm}
\facility{CARMENES/3.5 m telescope on Calar Alto.}


\software{\texttt{batman} \citep{kreidberg2015}, \texttt{corner} \citep{foremanmackey2016}, \texttt{emcee} \citep{foremanmackey2013}, \texttt{FastChem} \citep{stock2018}, \texttt{HELIOS-K} \citep{grimm2021}, \texttt{matplotlib} \citep{hunter2007}, \texttt{molecfit} \citep{smette2015}, \texttt{numpy} \citep{numpy2020}, \texttt{p-winds} \citep{dossantos2022}, \texttt{Spectroscopy Made Easy} \citep[][]{piskunov2017}.}




\bibliography{sample631}{}
\bibliographystyle{aasjournal}



\end{document}